\pdfoutput=1
\documentclass[preprint,12pt,authoryear]{elsarticle}

\usepackage{amssymb}
\usepackage{graphicx}
\usepackage{subfigure}
\usepackage{dcolumn}
\usepackage{bm}
\usepackage{natbib}
\setcitestyle{numbers,square}

\begin{document}

\title{High-Accuracy Approximation of Evolutionary Pairwise Games on Complex Networks}

\author[1]{Hongyu Wang}
\ead{wanghongyu1998@pku.edu.cn}

\author[1]{Aming Li\corref{cor1}}
\ead{liaming@pku.edu.cn}

\author[1]{Long Wang\corref{cor1}}
  
\ead{longwang@pku.edu.cn}

\cortext[cor1]{Corresponding authors}

\address[1]{Center for Systems and Control, College of Engineering, Peking University, 
Beijing, 100871, China}

\begin{abstract}

Previous studies have shown that the topological properties of a complex network, such as heterogeneity and average degree, affect the evolutionary game dynamics on it. However, traditional numerical simulations are usually time-consuming and demand a lot of computational resources. In this paper, we propose the method of dynamical approximate master equations (DAMEs) to accurately approximate the evolutionary outcomes on complex networks. We demonstrate that the accuracy of DAMEs supersedes previous standard pairwise approximation methods, and DAMEs require far fewer computational resources than traditional numerical simulations. We use prisoner's dilemma and snowdrift game on regular and scale-free networks to demonstrate the applicability of DAMEs. Overall, our method facilitates the investigation of evolutionary dynamics on a broad range of complex networks, and provides new insights into the puzzle of cooperation.
\end{abstract}

\maketitle

\clearpage
\section{\label{sec:level1} Introduction }

Many levels of biological organization, from single-celled organisms to human society, are based on cooperation \cite{Axelrod81}. However, in the context of Darwinian evolution, natural selection favors defectors over cooperators. Evolutionary game theory is a general mathematical framework for studying the  cooperation between unrelated individuals \cite{Maynard1982,SIGMUND03,Hauert06,Nowak06,Traulsen06,allen2014}. As metaphors for studying the evolution of cooperation, pairwise games such as prisoner's dilemma (PD) and snowdrift game (SG) have been widely adopted by researchers from different backgrounds \cite{Axelrod81,Nowak92,Hauert04,Santos05,Hauert05}. In infinitely well-mixed populations,  evolution under replicator dynamics leads to a stable fraction of cooperators for SG but to the complete extinction of cooperators in PD \cite{Schuster1983,Hofbauer1998}.

To better understand the emergence of cooperation in more realistic situations, the importance of studying the behavior of individuals with population structure should be highlighted. Graph theory provides a convenient framework to describe the population structure for studying the evolution of cooperation \cite{Santos06,Ohtsuki06,Szabo07,FuF07,FuF07heterogeneous,CHEN2007,FuF2008,FuF2009,WuB10,Gomez12,Perc13,su2019,LiA20}, where the vertices of a graph represent players and the edges define the network of contacts between players. Researchers found that when individuals interacted only with their neighbors, the fraction of cooperators in both PD and SG differs from that in well-mixed populations \cite{Nowak92,Hauert04,Santos06,Santos2006PNAS,Szabo07,Perc08,Roca09}. Regular graphs ignore the uniqueness of individuals, that is, different individuals may have different numbers of neighbors they interact with. Scale-free (SF) networks, whose vertex connectivities follow a power-law distribution, are often used to represent more realistic heterogeneous networks \cite{BA1999}. Santos and Pacheco found that the equilibrium frequencies of cooperators are higher when playing the PD and the SG on SF networks than when playing them on regular networks \cite{Santos05}. They attributed this phenomenon to the generation rules of the Barab\'{a}si-Albert model, that is, the vertices in the network are added sequentially, and the newly added vertice is more likely to connect to vertices with higher degrees. However, the lack of more in-depth studies, particularly a theoretical explanation, makes this phenomenon challenging to understand. Ohtsuki et al. proved that natural selection favors cooperation if the benefit of altruistic behavior, divided by the costs, exceeds the average number of neighbors \cite{Ohtsuki06}. Recent works explore general formulations of fixation probabilities for pairwise games under weak selection that apply to graph-structured populations \cite{allen2017,McAvoy21,allen2021,SuQ2022}. In this paper, we propose the dynamical approximate master equations (DAMEs) to describe evolutionary game dynamics on complex networks, starting from a theoretical explanation of how the heterogeneity of networks affects the evolutionary dynamics and finally leads to the prevalence of cooperation on SF networks.

Indeed, beyond numerical simulations, we need a theoretical method to study the evolutionary behavior of various nodes on complex networks. The most commonly used method for binary-state dynamics on complex networks is the mean-field (MF) theory \cite{MF5,MF2,MF4,MF3,MF1}. The pairwise approximation (PA) theory suggested by Dickman improves the accuracy of MF theory \cite{PA1,Levin1996,PA2,Taylor12,Mata14} and has been applied to the study of evolutionary dynamics on complex networks \cite{traulsen2007,Fu2009,WangX2012,SuiX2015,cnx018}. The approximate master equations (AMEs) have achieved high accuracy beyond the PA level in the binary-state dynamics on complex networks \cite{AMEJTB,PRL11,PRX13,peralta2018}, but the transition probabilities between two states in the AME system are static (time-invariant). In evolutionary games on complex networks, the transition probabilities between states of each node depend on the difference between the node and its neighbors' payoff, which is time-variant. Here we incorporate the transition probabilities in the time-variant form in our DAMEs, which can accurately approximate evolutionary game dynamics and predict the equilibrium frequencies of cooperators in a short time.

\section{\label{sec:level2}  Evolutionary pairwise games on complex networks}
We consider a population captured by the complex network with $N$ nodes. Each player in the population can be in two states, cooperation or defection. The degree of a node represents the number of edges between the node and its neighbors. Degree distribution $P(k)=N_{k} / N$ represents the probability that the degree of a randomly selected node in the network is $k$, where $N_{k}$ refers to the number of nodes with degree $k$. Here we assume that the network is generated by the configuration model with fixed degree distribution  $P(k)$, which does not have degree correlations \cite{CM1}.

In both PD and SG, two players have to decide whether to cooperate or defect in each round. Both players receive $R$ when they cooperate with each other and $P$ when they defect. A defector exploiting a cooperator receives $T$, and the cooperator receives $S$. Following common practice, researchers usually adjust the game to rely on a single parameter. For the PD, we have $T > R > P > S$. Considering that $T + S < 2R$, we make $ 2 > T = b > 1$, $R = 1$, and $P = S= 0$, leaving the advantage of defectors $b$ be the single parameter. We have tested that if $S = -\varepsilon<0$ $(\varepsilon \ll 1)$ is set to satisfy $S < P$, the result will not change. For the SG, we have $T > R > S > P$. Considering that $T + S = 2R$, we make $ T = \beta > 1$, $R = \beta - 1/2$, $S = \beta - 1$, and $ P= 0$, such that the cost-to-benefit ratio can be written as $r = 1 / (2\beta -1)$.

Evolution is carried out by implementing the finite population analog of replicator dynamics through the following transition probabilities: In each generation, all individuals play a single-round game with all of their neighbors and accumulate the payoffs. Whenever an individual $i$ desires to update the strategy, one of its neighbor $j$ will be drawn from its $k_{i}$ neighbors. The probability that the individual $i$ copies the strategy of individual $j$ is given by the Fermi function
\begin{equation}
P_{s_{i} \rightarrow s{j}}=\phi(\pi_{i},\pi_{j})=\frac{1}{1+e^{\alpha\left(\pi_{i}-\pi_{j}\right)}}\nonumber
\end{equation}
where $\alpha \in[0, \infty)$ denotes the intensity of selection. $\alpha \rightarrow 0$ leads to the random drift and $\alpha \rightarrow \infty$ leads to the deterministic imitation dynamics.

\section{\label{sec:level3}  Dynamical approximate master equations}

Define $C_{k, m}(t)$ ($D_{k, m}(t)$) as  the fraction of $k$-degree nodes that are cooperators (defectors) at time $t$ and have $m$ defector neighbors. The DAMEs consist of  $M=\sum_{k, m}2 =(1+k_{max}-k_{min})(2+k_{max}+k_{min})$ variables.   Then the fraction of cooperators of $k$-degree nodes at time $t$ is given by
\begin{equation}
 \rho_{k}(t)=\sum_{m=0}^{k} C_{k, m}(t),\nonumber
\end{equation}
and the fraction of cooperators in the whole network is 
\begin{equation}
\rho(t)=\sum_{k} P(k) \rho_{k}(t). 
\nonumber
\end{equation}

By assuming that the initial cooperators are randomly selected with a fraction $\rho(0)$, the initial conditions are  
\begin{eqnarray}
C_{k, m}(0)=\rho(0) B_{k, m}(1-\rho(0)), 
\\
D_{k, m}(0)=(1-\rho(0)) B_{k, m}(1-\rho(0)), 
\end{eqnarray}
where $B_{k, m}(q)=\left(\begin{array}{c}k \\ m\end{array}\right) q^{m}(1-q)^{k-m}$ is the binomial factor.

The approximate master equations for the evolution of $C_{k, m}(t)$ and $D_{k, m}(t)$ are
\begin{eqnarray}
\frac{d C_{k, m}(t)}{d t} =&&-P_{C_{k, m} \rightarrow D_{k, m}} C_{k, m}(t) \frac{m}{k} \nonumber \\
&&+P_{D_{k, m} \rightarrow C_{k, m}} D_{k, m}(t) \frac{(k-m)}{k} \nonumber \\
&&-P_{C_{k, m} \rightarrow C_{k, m+1}}(k-m) \beta^{C}C_{k, m}(t)  \nonumber\\
&&+P_{C_{k, m-1} \rightarrow C_{k, m}}(k-m+1) \beta^{C}C_{k, m-1}(t)  \nonumber\\
&&-P_{C_{k, m} \rightarrow C_{k, m-1}} m\gamma^{C} C_{k, m}(t)  \nonumber\\
&&+P_{C_{k, m+1} \rightarrow C_{k, m}}(m+1)\gamma^{C} C_{k, m+1}(t) ,
\label{eqa:C}
\end{eqnarray}
\begin{eqnarray}
\frac{d D_{k, m}(t)}{d t}=&&-P_{D_{k, m} \rightarrow C_{k, m}} D_{k, m}(t) \frac{(k-m)}{k} \nonumber \\
&&+P_{C_{k, m} \rightarrow D_{k, m}} C_{k, m}(t) \frac{m}{k} \nonumber \\
&&-P_{D_{k, m} \rightarrow D_{k, m+1}}(k-m) \beta^{D}D_{k, m}(t)  \nonumber \\
&&+P_{D_{k, m-1} \rightarrow D_{k, m}}(k-m+1) \beta^{D}D_{k, m-1}(t)  \nonumber \\
&&-P_{D_{k, m} \rightarrow D_{k, m-1}} m \gamma^{D}D_{k, m}(t)  \nonumber \\
&&+P_{D_{k, m+1} \rightarrow D_{k, m}}(m+1) \gamma^{D}D_{k, m+1}(t) .
\end{eqnarray}

\begin{figure}[ht] 
  \centering
  \includegraphics[scale=0.9]{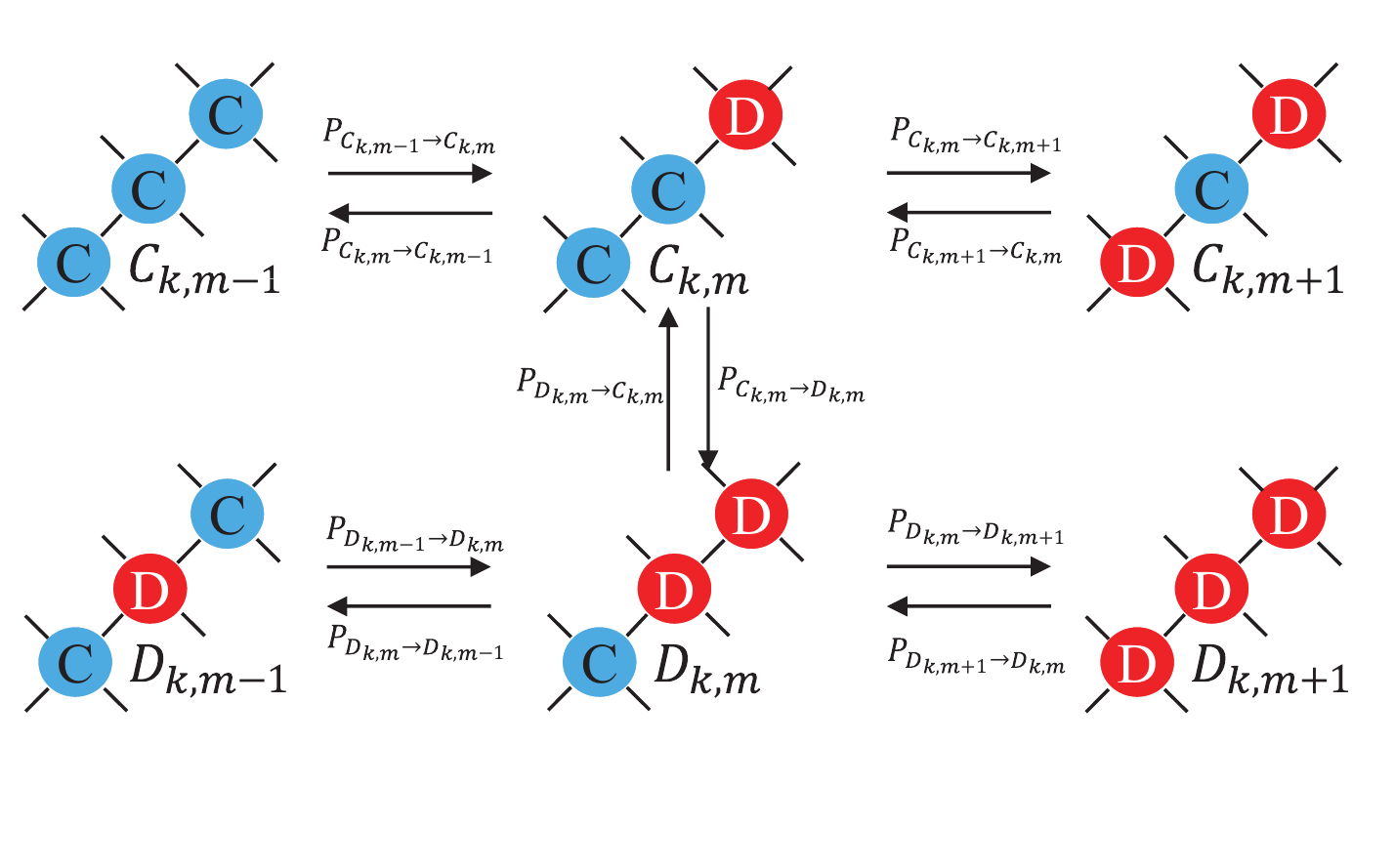}   
   
  \caption{ \label{fig:0} Schematic representation of the meaning of each variable in DAMEs. Define $C_{k, m}(t)$ ($D_{k, m}(t)$) as  the fraction of $k$-degree nodes that are cooperators (defectors) at time $t$ and have $m$ neighboring defectors. $C_{k, m}(t)$ and $D_{k, m}(t)$ change at each time step due to the node itself or its neighbors' updating strategies. For example, the coefficient $P_{C_{k, m} \rightarrow D_{k, m}}$ is defined as the transition probability that a $k$-degree cooperator, which has $m$ defector neighbors at time $t$, changes its strategy to defection by time $t+d t$. A node may update its strategy only when it plays with nodes that adopt the different strategy, and the probability that a $C_{k, m}(t)$ node plays with its defector neighbors is $\frac{m}{k}$.}
\end{figure}

The coefficient $P_{C_{k, m} \rightarrow D_{k, m}}$ is defined as the transition probability that a $k$-degree cooperator, which has $m$ defector neighbors at time $t$, changes its strategy to defection by time $t+d t$, where $dt$ is an infinitesimally small time interval. Similarly, $P_{D_{k, m} \rightarrow C_{k, m}}$ is the transition probability that a $k$-degree defector, which has $m$ defector neighbors at time $t$, changes its strategy to cooperation by time $t+d t$.

The coefficient $P_{C_{k, m} \rightarrow C_{k, m+1}}$ is defined as the transition probability that a cooperator, which has $k$ neighbors and $m$ of them are defectors at time $t$, changes its state into $C_{k, m+1}$ by time $t+d t$, which means one of its cooperator neighbors becomes a defector. The coefficient $\beta^{C}$ is defined as the probability that the randomly selected neighbor is a defector when a cooperator's cooperator neighbor updates its strategy. The coefficient $P_{C_{k, m} \rightarrow C_{k, m-1}}$ is defined as the transition probability that a cooperator, which has $k$ neighbors and $m$ of them are defectors at time $t$, changes its state into $C_{k, m-1}$ by time $t+d t$, which means one of its defector neighbors becomes a cooperator. The coefficient $\gamma^{C}$ is defined as the probability that the randomly selected neighbor is a cooperator when a cooperator's defector neighbor updates its strategy. $P_{C_{k, m-1} \rightarrow C_{k, m}}$, $P_{C_{k, m} \rightarrow C_{k, m-1}}$, $P_{C_{k, m+1} \rightarrow C_{k, m}}$, $P_{D_{k, m} \rightarrow D_{k, m+1}}$, $P_{D_{k, m-1} \rightarrow D_{k, m}}$, $P_{D_{k, m} \rightarrow D_{k, m-1}}$ and $P_{D_{k, m+1} \rightarrow D_{k, m}}$, $\beta^{D}$, $\gamma^{D}$ are defined in the same way. The mathematical expressions of these variables are given somewhere below.

In order to compute these transition probabilities, we can directly calculate the payoff of each class of nodes and use the limited information in the DAMEs system to estimate the payoff of each class of nodes' first-order neighbors and second-order neighbors. The approximation in this method is that we assume that the neighbor configuration of each node is the product of independent single event probability. That is, the network under study should have no degree correlations, which will lead to inaccurate results of our model under some extreme networks (such as social networks and e-mail networks).

Step 1: Compute the number of cooperator neighbors and defector neighbors of nodes with different degree.

The number of defector neighbors of a cooperator with degree $k_{i}$ can be computed by
\begin{equation}
N_{C-D}(k_{i})=\frac{\sum_{m=0}^{k_{i}}m C_{k_{i}, m}} {\sum_{m=0}^{k_{i}}  C_{k_{i}, m}} .\nonumber
\end{equation}
$N_{C-C}(k_{i})$, $N_{D-C}(k_{i})$, $N_{D-D}(k_{i})$ are defined similarly.

Step 2: Compute the degree distribution and payoffs of each class of nodes' first-order neighbors.

We can compute the degree distribution of cooperators which have at least a cooperator neighbor as
\begin{equation}
P_{k C-C}(k_{i})=\frac{P(k_{i}) \sum_{m=0}^{k_{i}-1} C_{k_{i}, m}}{\sum_{k} P(k)\left(\sum_{m=0}^{k-1} C_{k, m}\right)},\nonumber
\end{equation}
$k_{i}\in [k_{min},k_{max}]$. The reason that the integral has an upper bound of $k_{i}-1$ is that a node with degree $k_{i}$ can have at most $k_{i}-1$ defector neighbors to ensure that it has at least one cooperator neighbor. The probability of a node connecting with nodes with degree $k_{i}$ depends on the ratio of the number of its edges to the number of all edges. Thus the degree distribution of first-order cooperator neighbors of a cooperator can be computed as:
\begin{equation}
P_{C-C}(k_{i})=\frac{k_{i} P_{k C-C}(k_{i})}{\sum_{k} k P_{k C-C}(k)}.\nonumber
\end{equation}

The payoffs of a cooperator's first-order cooperator neighbors depend on the number of their cooperator neighbors and defector neighbors.
\begin{equation}
\pi_{C-C}(k_{i})=\frac{\sum_{m=0}^{k_{i-1}}(k_{i}-m) C_{k_{i}, m}} {\sum_{m=0}^{k_{i-1}}  C_{k_{i}, m}} \cdot R + \frac{\sum_{m=0}^{k_{i-1}}m C_{k_{i}, m}} {\sum_{m=0}^{k_{i-1}}  C_{k_{i}, m}} \cdot S.\nonumber
\end{equation}

The fraction of defector neighbors of a cooperator's cooperator neighbor can be computed by

\begin{equation}
\beta^{C}=\frac{\sum_{k} P(k)m \sum_{m=0}^{k-1} C_{k, m}}{\sum_{k} P(k)k\left(\sum_{m=0}^{k-1} C_{k, m}\right)}.\nonumber
\end{equation}
$\beta^{D}$, $\gamma^{C}$, $\gamma^{D}$ can be computed similarly.

Step 3: Compute $P_{C_{k, m} \rightarrow D_{k, m}}$ and $P_{D_{k, m} \rightarrow C_{k, m}}$, in which case the focal players are selected to update strategies.

When the focal players is a defector, its payoff is
\begin{equation}
\pi_{D}(k,m)=(k-m)  T+m  P.\nonumber
\end{equation}

Using the conclusion of Step 2, we can compute

\begin{equation}
P_{kC-D} (k_{i})=\frac{P(k_{i}) \sum_{m=1}^{k_{i}} C_{k_{i}, m}}{\sum_{k} P(k)\left(\sum_{m=1}^{k_{i}} C_{k, m}\right)},\nonumber
\end{equation}
\begin{equation}
P_{D-C}(k_{i})=\frac{k_{i} P_{k C-D}(k_{i})}{\sum_{k} k P_{k C-D}(k)},\nonumber
\end{equation}
\begin{equation}
\pi_{D-C}(k)=\frac{\sum_{m=1}^{k_{i}}(k_{i}-m) C_{k_{i}, m}} {\sum_{m=1}^{k_{i}}  C_{k_{i}, m}} \cdot R+\frac{\sum_{m=1}^{k_{i}} m C_{k_{i}, m}} {\sum_{m=1}^{k_{i}} C_{k_{i}, m}} \cdot S.\nonumber
\end{equation}

Thus the transition probability $P_{D_{k, m} \rightarrow C_{k, m}}$ can be computed as
\begin{equation}
P_{D_{k, m} \rightarrow C_{k, m}}=\sum_{k_{i}=k_{min}}^{k_{max}} P_{D-C}(k_{i})\phi(\pi_{D}(k,m),\pi_{D-C}(k_{i})).\nonumber
\end{equation}

Because the probability that a randomly selected neighbor is a cooperator is $(k-m)/k$, we multiply this coefficient in the second line in Eq.~(\ref{eqa:C}). $P_{C_{k, m} \rightarrow D_{k, m}}$ can be computed in the same way.

Step 4: Compute other transition probabilities, in which case the neighbors of the focal players are selected to update strategies.

$P_{C_{k, m} \rightarrow C_{k, m+1}}$ is defined as the probability that one of the cooperator neighbors of a focal player $A$ in class $C_{k, m}$ becomes a defector by time $t+d t$. $A$ is a cooperator, so this  first-order cooperator neighbor must be connected with a second-order defector neighbor. We can compute $A$'s first-order cooperator neighbors' degree distribution $P_{C-C}(k_{i})$ and payoff $\pi_{C-C}(k_{i})$ by the method in Step 2. We can also compute $A$'s second-order defector neighbors' degree distribution $P_{C-D}(k_{j})$ and payoff $\pi_{C-D}(k_{j})$. Thus $P_{C_{k, m} \rightarrow C_{k, m+1}}$ is given by
{
\begin{eqnarray}
P_{C_{k, m} \rightarrow C_{k, m+1}}= && \sum_{k_{i}} \sum_{k_{j}} P_{C-C}(k_{i}) P_{C-D}(k_{j}) \nonumber  
\\
&& \times \phi(\pi_{C-C}(k_{i}),\pi_{C-D}(k_{j})). 
\end{eqnarray}
}

$P_{C_{k, m-1} \rightarrow C_{k, m}}$, $P_{D_{k, m} \rightarrow D_{k, m-1}}$ and $P_{D_{k, m+1} \rightarrow D_{k, m}}$ can be computed similarly because in these cases first-order neighbors can only update their strategy by imitating the strategies of second-order neighbors.

$P_{C_{k, m} \rightarrow C_{k, m-1}}$ is defined as the probability that one of the defector neighbors of a focal player $B$ in class $C_{k, m}$ becomes a cooperator by time $t+d t$. We must consider that the first-order defector neighbors have probabilities to imitate $B$'s strategy. Similarly, we can compute the first-order defector neighbors' degree distribution $P_{C-D}(k_{i})$ and their payoff $\pi_{C-D}(k_{i})$ and the second-order cooperator neighbors' degree distribution $P_{D-C}(k_{j})$ and payoff $\pi_{D-C}(k_{j})$. The number of cooperator neighbors of first-order defector neighbors(with degree $k_{i}$) can be computed by the method in Step 1:
\begin{equation}
N_{D-C}(k_{i})=\frac{\sum_{m=0}^{k_{i}-1}(k_{i}-m) D_{k_{i}, m}} {\sum_{m=0}^{k_{i}-1}  D_{k_{i}, m}} .\nonumber
\end{equation}
Thus first-order defector neighbors with degree $k_{i}$ have a probability of $1/N_{D-C}(k_{i})$ to imitate $B$'s strategy, and have a probability of $[N_{D-C}(k_{i})-1]/N_{D-C}(k_{i})$ to imitate a second-order cooperator node' strategy. Therefore $P_{C_{k, m} \rightarrow C_{k, m-1}}$ is computed by
\begin{eqnarray}
&&P_{C_{k, m} \rightarrow C_{k, m-1}}(k, m)=  
\sum_{k_{i}} P_{C-D}(k_{i}) \left(\frac{1}{N_{D-C}(k_{i})}\right.  \nonumber \\
&& \left.  \times \phi(\pi_{C-D}(k_{i}),\pi_{C}(k, m)) + \frac{N_{D-C}(k_{i})-1}{N_{D-C}(k_{i})} \right. \nonumber 
\\
&& \left. \times \sum_{k_{j}} P_{D-C}(k_{j})  \phi(\pi_{C-D}(k_{i}),\pi_{D-C}(k_{j}))\right).
\end{eqnarray}
 $P_{C_{k, m+1} \rightarrow C_{k, m}}$, $P_{D_{k, m} \rightarrow D_{k, m+1}}$ and $P_{D_{k, m-1} \rightarrow D_{k, m}}$ can be computed similarly.

\section{\label{sec:level4}  Simulations of evolutionary pairwise games on networks}
To test the accuracy of the approximations, we will now compare the results of the theory explained in the previous section to the numerical simulations. We consider the simulated dynamics on different types of networks with $N = 10000$ nodes. We consider two kinds of networks with different degree distributions: regular ring networks and scale-free networks. For scale-free networks, the degree distribution $P(k) \sim k^{-\gamma}, 2 \leq \gamma \leq 3$ obeys the power law. For convenience, we use the configuration model to generate scale-free networks with degree fixed distribution  $P(k)$, which is given by the Barab\'{a}si-Albert model. The initial fraction of cooperators is $\rho = 50\%$. Our simulation results are the average of more than 1000 independent simulations. PAs and DAMEs used networks with the same degree distribution $P(k)$ as simulations.

\subsection{Equilibrium frequencies of cooperators}

We first consider the influence of network structure and average connectivity $z$ on the evolution of cooperation. The role of average connectivity in cooperation has been examined by several previous works \cite{Santos05,Santos06}. We calculate the average equilibrium frequencies of cooperators of different models in the steady state. We will focus on the differences between simulation results and theoretical results given by DAMEs and the PA method. 

\begin{figure}[h] 
  \centering
  \includegraphics[scale=0.5]{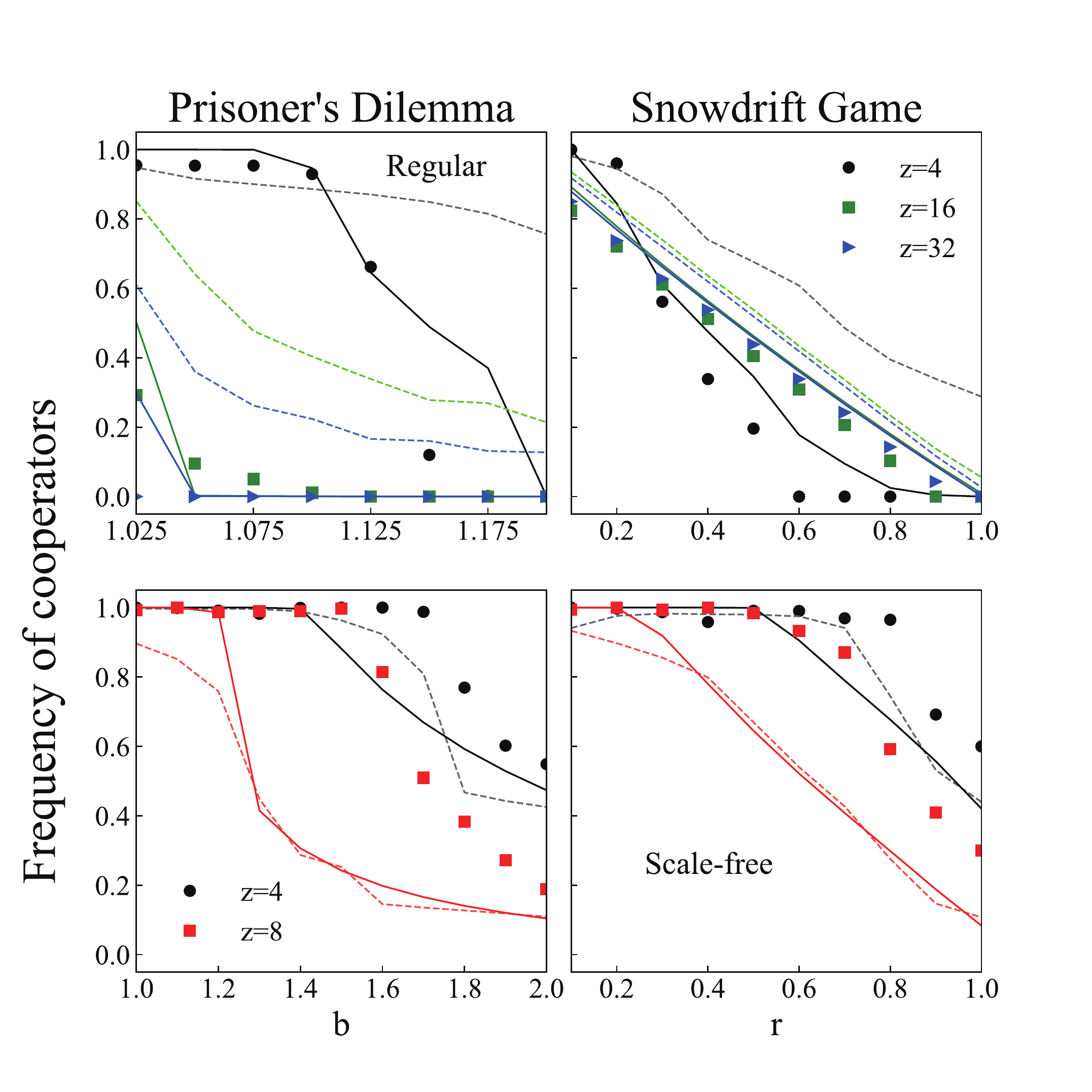}   
 
  \caption{ \label{fig:1}   The equilibrium frequencies of cooperators on different types of networks. Results are shown as functions of advantage of defectors $b$ for the PD (left panels) and cost-to-benefit ratio $r$ for the SG (right panels). Results for regular ring networks are shown on top panels and for scale-free networks on lower panels. Markers, dashed lines, and solid lines indicate the results of simulations, PAs, and DAMEs. Different average connectivity $z$ is distinguished by different colors.} 
\end{figure}

Firstly, we consider the evolutionary games on regular ring networks. As the average connectivity $z$ of the network increases, the equilibrium frequencies of cooperators decrease rapidly for the PD (top left panel of Fig.~\ref{fig:1}). For the SG (top right panel of Fig.~\ref{fig:1}), the gradual increase of $z$ results in the equilibrium frequencies of cooperators gradually approaching $1-r$, which is consistent with the result concluded by the replicator equation in well-mixed populations. In some parameter spaces, the equilibrium frequencies of cooperators for the SG on regular ring networks are significantly lower than $1-r$. This phenomenon is well described by DAMEs, but PAs do not give correct predictions. We show that DAMEs clearly give a better approximation to evolutionary dynamics than the PA method for both cases on regular ring networks. It is easy to find that when $z$ is large enough, the properties of regular ring networks are similar to well-mixed populations. We also find that the PA method will give a higher equilibrium frequency of cooperators than the simulation value in most cases. We speculate that this is because the PA method classifies nodes only by their degree and can not capture the behavior differences between nodes with different types of neighbors.

Aside from regular ring networks, we now turn our attention to the study of evolutionary dynamics on scale-free networks. Different from results obtained by Santos and Pacheco on SF networks \cite{Santos06}, for both the PD and the SG, the equilibrium frequencies of cooperators decreased significantly when the average connectivity changed from 4 to 8 (lower panel of Fig.~\ref{fig:1}), which is the same as that observed in regular ring networks. In scale-free networks, nodes with cooperative strategies often form clusters of various sizes. These clusters will protect internal cooperators from being invaded by defectors, even if the payoffs of these cooperators are not high enough. DAMEs focus on characterizing the state of each type of node and its first-order neighbors, resulting in a lower prediction of the equilibrium frequencies of cooperators.  Although there are differences between the results of DAMEs, PAs, and simulation results, they show the same trend variation. With the increase of average connectivity $z$ and the intensity of social dilemma, the equilibrium frequencies of cooperators gradually decrease.

\subsection{The behavior of various nodes in SF networks}

In order to study how scale-free networks promote the emergence of cooperation in evolutionary games, we must pay attention to the main difference between scale-free networks, well-mixed populations, regular networks, and random networks, that is, the heterogeneity of networks. Several previous works have investigated the key role of network topology in the evolution of cooperation \cite{Santos05,Santos06,FuF07,FuF07heterogeneous,roca2009,LiA20}. 
Specifically, we focus on the frequencies of cooperators with different degrees and how they behave during the process of evolution, and we study the evolutionary dynamics of the PD with $b = 1.4$ on scale-free networks. In contrast, when we replace scale-free networks with regular-ring networks, the equilibrium frequencies of cooperators become 0. We divide nodes into four types according to their degrees: $Central$ nodes with degree $k_{i} \geqslant 10$ (top 5.5\%), $Large$ nodes with degree $10 > k_{i} \geqslant 5$ (5.5\% - 20\%), $Middle$ nodes with degree $5 > k_{i} \geqslant 3$ (20\% - 50\%), and $Small$: nodes  with degree $k_{i} < 3$ (50\% - 100\%).

\begin{figure}[h] 
  \centering
  \includegraphics[scale=0.5]{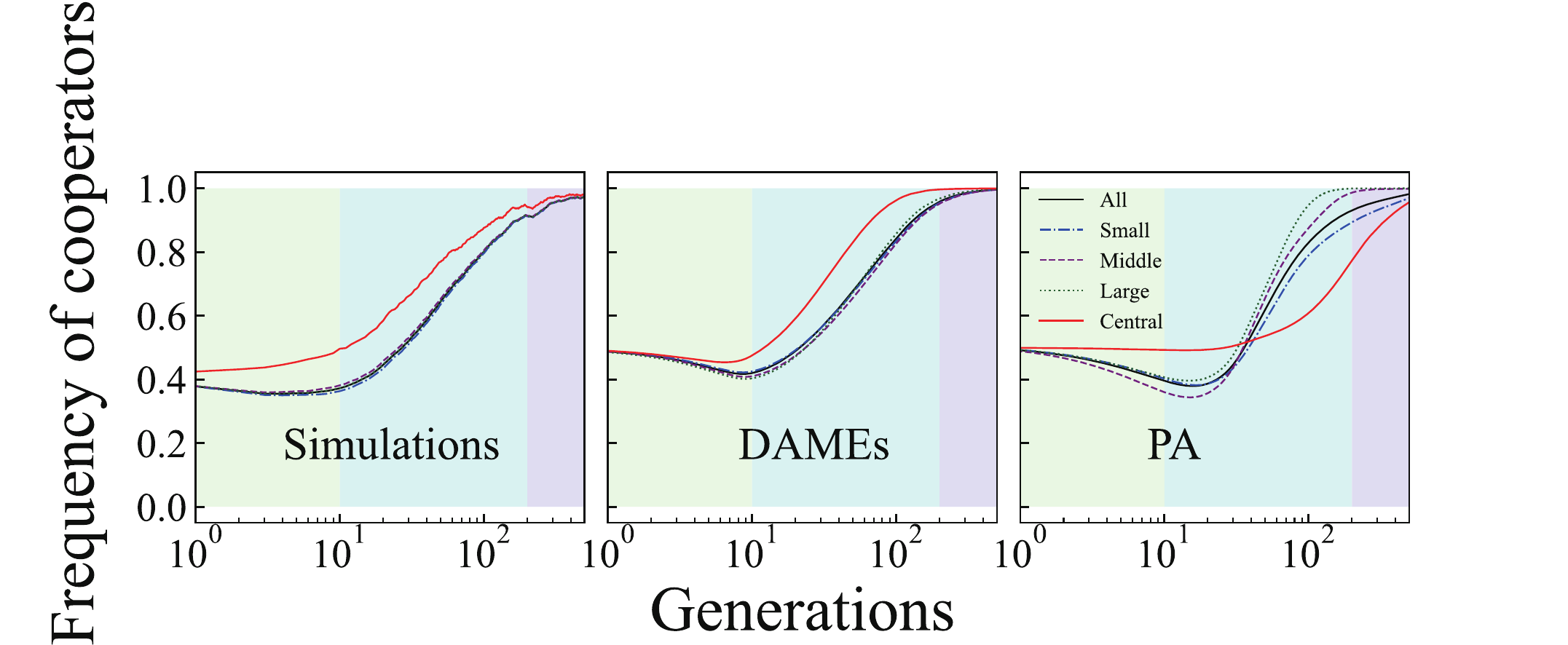}    

  \caption{\label{fig:2} Evolutionary dynamics of the prisoner's dilemma with $b = 1.4$ on scale-free networks with 10,000 nodes and an average connectivity $z = 4$. Results are shown as functions of generations. The initial fraction of cooperators is $\rho = 50\%$. Nodes with different degrees are distinguished by different colors.}  
\end{figure}

At about the tenth generation, we find that the frequency of cooperators in the networks reaches the lowest level, and then the frequency of cooperators slowly rises. After about 500 generations, the vast majority of nodes in the networks are cooperators. A remarkable feature of evolutionary dynamics is that the frequency of cooperators of the $Central$ nodes is significantly higher than those of other nodes, and the frequencies of cooperators of all other nodes are almost the same at all generations. DAMEs (middle panel of Fig.~\ref{fig:1}) capture these characteristics of evolutionary dynamics quantitatively, but PAs (right panel of Fig.~\ref{fig:1}) can not even capture these features qualitatively. By analyzing $C_{k, m}(t)$ and $D_{k, m}(t)$ in DAME approximations, we can divide the evolution of cooperation in this situation into three phases: 

The evolution begins with the alienation phase. In the beginning, the cooperators are randomly distributed in the network. Since the average payoff of the defectors is 1.4 times higher than the cooperators', the frequency of cooperators decreases, but the $Central$ nodes are less affected due to their higher payoffs. Then, $Central$ nodes with higher payoffs gradually propagate their own strategies. $Central$ cooperators gradually turn their neighbors into cooperators while $Central$ defectors gradually turn their neighbors into defectors. At this phase, clusters of nodes of the same type appear. Clusters generally have several $Central$ nodes and $Large$ nodes.

Then the rising phase appears. $Central$ cooperators get higher payoffs and become more stable, while $Central$ defectors get lower payoffs. When their payoffs are low enough, there is a high probability of learning the strategies from their cooperator neighbors. Through the above process, the frequency of cooperators of the $Central$ node gradually increases. The $Central$ nodes have many neighbors, which means that as the evolution progresses, the probability of other nodes contacting high-payoff cooperators increases. Nodes that contact the $Central$ nodes gradually become cooperators, bringing the frequency of cooperators in the network increases. Increased frequency of cooperators makes it easier for defectors to contact higher-payoff cooperators. 

The last is the balance phase. After a long period of evolution, the vast majority of nodes in the network become cooperators (when $b = 1.4$). In most cases, there are still some defectors in the network, and the transition between cooperators and defectors reaches a balance. That is, the fraction of cooperators in the network remains stable. Even at this phase, the frequency of cooperators of the $Central$ nodes is still significantly higher than that of other nodes.

In regular ring networks, there are no $Central$ nodes that provide leadership, and all cooperators turn into defectors shortly because their payoffs are much lower than those of the defectors. The analysis of nodes' behavior in the simulations confirms the above process. Compared with traditional numerical simulations that consume a lot of computing resources (proportional to network size ${N}^2$ and usually require a large number of repeated experiments), the DAMEs can approximate evolutionary dynamics with high accuracy in a very short time (proportional to ${k^2_{max}}$). For that $k_{\max }\propto N^{\frac{1}{\gamma-1}}$ in scale-free networks and $k_{\max } \propto \ln N$ in regular networks, it is easy to find that DAMEs save a lot of computational resources.   The DAMEs provide a convenient theoretical analysis framework, which can accurately demonstrate the critical role of a few highly connected nodes (hubs) in SF networks. Moreover, it can easily capture the evolutionary behavior of various nodes in complex networks.

\subsection{The behavior of edges in complex networks}

\begin{figure}[h] 
  \centering
  \includegraphics[scale=0.5]{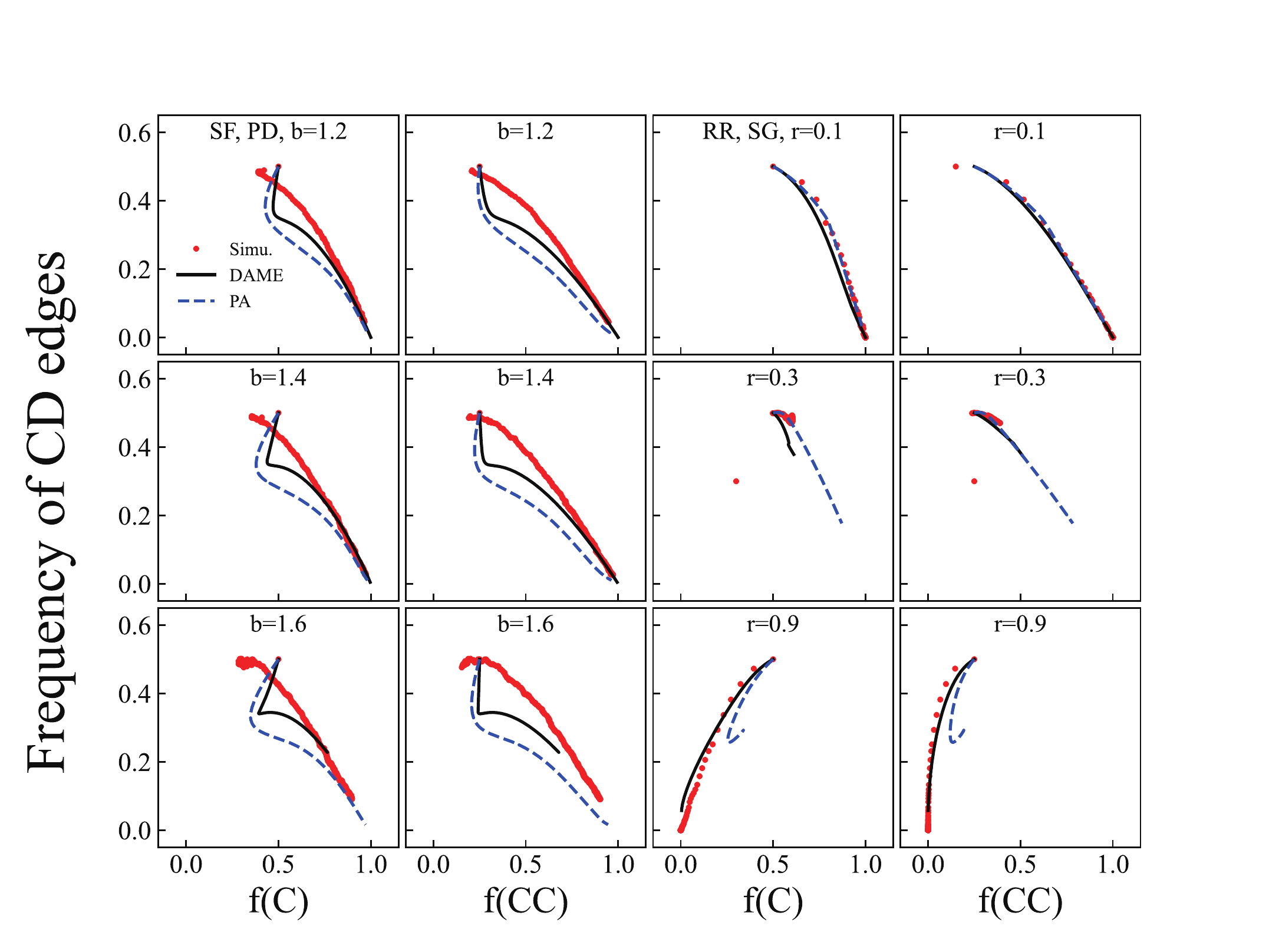}   
  \caption{ \label{fig:3} The trajectories of $(C, CD)$ and $(CC, CD)$ coordinates. Markers, dashed lines, and solid lines indicate simulation results, PA results, and DAME results. All cases have the same average connectivity $z = 4$. We can find that DAMEs have better overall accuracy than PAs for the PD on SF networks. For the SG on RR networks, both DAMEs and PAs perform well when $r=0.1$. DAMEs accurately describe the evolution dynamics when $r=0.3$ or $r=0.9$, but PAs give wrong results. $f(C)$ is the frequency of cooperators and $f(CC)$ is the frequency of $CC$ edges.}  
\end{figure}

Having explored the equilibrium frequencies of cooperators and the behavior of nodes during evolution, we next explore the behavior of edges in the network as the network dynamics. The initial fraction of cooperators, $CC$ edges, and $CD$ edges in each case are $50\%$, $25\%$, and $50\%$. As evolution progresses, these fractions change, and the trajectories of $(C, CD)$ and $(CC, CD)$ coordinates are shown in Fig.~\ref{fig:3}. To demonstrate the applicability of DAMEs, we study the evolutionary dynamics of the prisoner's dilemma on scale-free networks with different $b$ and that of the snowdrift game on regular ring networks with different $r$.

We first demonstrate the evolution of the $CC$ and $CD$ edges of the prisoner's dilemma on scale-free networks (left two columns of Fig.~\ref{fig:3}). Both PAs and DAMEs capture the rapid decline of the fraction of cooperators during the alienation phase, but both approximation methods overestimate the rate of decline in the fraction of $CD$ edges. We also find that both approximations, in general, can qualitatively give the change in the fraction of the edges, and the DAME approximations have better overall accuracy.

We further show the evolution of the $CC$ and $CD$ edges of the snowdrift game on regular ring networks (right two columns of Fig.~\ref{fig:3}). For the case of $r=0.1$, PAs and DAMEs achieve equally high approximation accuracy. PAs fail to give correct evolutionary dynamics as the cost-to-benefit ratio $r$ increases, but DAMEs still maintain pretty high accuracy.

\section{\label{sec:level5}  Discussion and Conclusion}

In this paper, we have proposed the method DAMEs to theoretically predict the evolutionary game dynamics on complex networks. In particular, we apply it to study how network heterogeneity influences the evolutionary trajectories. On strongly heterogeneous networks such as scale-free networks, a small proportion of nodes are more highly connected than the majority, while the connectivity of nodes tends to be similar as the decreasing of network heterogeneity. Our results help to understand why network heterogeneity acts as a cooperation-promotor: Hubs are more likely to propagate their own strategies than the less influential nodes. Hubs adopting the cooperative strategy become more stable during evolution. Hubs adopting the defecting strategy reduce their payoff while propagating their own strategies, so that they are more likely to learn the cooperative strategy from their neighbors. Hubs will gradually become cooperators and propagate their own strategies, promoting the evolution of cooperation in the network. Furthermore, our findings also inspire a few possibilities to enhance the establishment of cooperation by network surgery or connection modification --- reconnect the edges so that the network has multiple, evenly distributed hubs with moderate influence, or set a suitable cut-off limit for the maximum connectivity of the nodes during network generation. The main objective of these solutions is to increase the proportion of hubs in the network.

By comparing the evolutionary dynamics given by numerical simulations, DAMEs, and PA methods with different spatial structures and payoff matrices, we demonstrated that the accuracy of DAMEs supersedes standard PA methods. MF methods, as a rather simple analytical approach, are often inaccurate on sparse networks due to the lack of dynamic correlations, which means that the state of a focal player's neighbors is assumed to be independent of the state of itself \cite{MF4}. PA methods consider dynamic correlations at a pairwise level but do not capture dynamical correlations beyond nearest neighbors \cite{Fu2009,PA2}. Classical AMEs achieve higher accuracy than MF methods and PA methods \cite{PRX13}, but the transition probabilities between two states in it are static and cannot be applied to evolutionary games. DAMEs consider the state of nodes and their first-order neighbors and have transition probabilities that depend on the difference between nodes' payoffs and estimated payoffs of the nodes' neighbors. 

To sum up, DAMEs, as a new tool for studying evolutionary games on complex networks, can better approximate evolutionary outcomes through large systems of differential equations. Using DAMEs for computing the equilibrium frequency of cooperators and the behavior of nodes and edges during evolution has been shown. Compared with traditional numerical methods, DAMEs may handle evolutionary games on large-scale networks with great efficiency and give reasonable evolutionary outcomes. DAMEs can also provide quick tests and helpful information for newly developed evolutionary game models. We expect that DAMEs will provide researchers with a viable alternative to computationally expensive and often time-consuming simulations.

\bibliographystyle{elsarticle-num} 
\bibliography{DAMEs}

\end{document}